\documentclass[fleqn,twoside]{article}
\usepackage{espcrc2}

\usepackage{graphicx}
\usepackage[figuresright]{rotating}
\usepackage{amsmath}

\begin{document}

\newcommand{\fives}{\Upsilon(5S)}
\newcommand{\fours}{\Upsilon(4S)}
\def\bbbar{B\bar{B}}
\def\bbstar{B^*\bar{B}}
\def\bstbst{B^*\bar{B}^*}
\def\bbbarpi{B\bar{B}\pi}
\def\bbstarpi{B^*\bar{B}\pi}
\def\bstbstpi{B^*\bar{B}^*\pi}
\def\bbbarpipi{B\bar{B}\pi\pi}
\def\bsbsbar{B^0_s\bar{B}^0_s}
\def\bsbsstar{B^{0*}_s\bar{B}^0_s}
\def\bsstbsst{B^{0*}_s\bar{B}^{0*}_s}
\def\ifb{{\rm fb^{-1}}}
\def\jpsi{J/\psi}
\def\de{\Delta E}
\def\mbc{m_{\rm bc}}
\def\ebeam{E_{\rm beam}}

\hyphenation{author another created financial paper re-commend-ed Post-Script}

\title{$\fives$: What has been learned and what can be learned}

\author{S. R. Blusk\address[MCSD]{Dept. of Physics, Syracuse University, Syracuse, NY 13244, USA}
        \thanks{We gratefully acknowledge 
the National Science Foundation for support of this work.}
(on behalf of the CLEO and Belle Collaborations)
}

\begin{abstract}
We present recent measurements of $B$ and $B^0_s$ production using data
collected on the $\fives$ resonance at CLEO and Belle. We also briefly discuss
what can be learned using sufficiently larger data samples in the future.
\vspace{1pc}
\end{abstract}

\maketitle


\section{INTRODUCTION}

The $\fives$ was discovered about 20 years ago in $e^+e^-$ collisions
at CLEO~\cite{cleo5s_85} and CUSB~\cite{cusb5s_85}. Its mass is 
$(10865\pm8)$~MeV, about 300 MeV 
higher 
than the $\fours$ resonance,
and its width of $(110\pm13)$~MeV is about five times larger
than the $\fours$ resonance. 
Its
production cross-section was measured
to be about 0.3~nb, about ten times lower than the $\fours$ cross-section.
Owing to its mass, the $\fives$ can decay to a variety of final states
containing $B$ and $B^0_s$ mesons. The accessible $B$-meson final states
include: $\bbbar$, $\bbstar$, $\bstbst$, $\bbbarpi$, $\bbstarpi$, $\bstbstpi$
and $\bbbarpipi$;
final states with $B^0_s$ are: $\bsbsbar$, $\bsbsstar$, and $\bsstbsst$.
Further study of the $\fives$ is of great interest because 
of it being a potentially  large source of $B^0_s$ mesons at an  $e^+e^-$ machine.
If suitably large
samples can be collected, their study 
would provide 
a complementary probe of 
CP violation and rare decays to LHCb. Studies involving
rare decays, particularly those with soft photons, and untagged time-dependent
and time-independent measurements, which provide access to 
the width difference between the $B^0_s$ mass eigenstates, 
$\Delta\Gamma_s$,
offer 
complementary information to measurements at hadron machines~\cite{nierste}.
Because of the small Lorentz boost at the $\fives$ and the large value of 
the mixing frequency, 
$\Delta m_s$~\cite{cdf}, time-dependent CP violation measurements in mixing 
are
not possible. 

Until recently, there existed only
weak limits on the production rate of $B^0_s$ at the $\fives$. Several models
attempt to describe the hadronic cross-section in the 10-11 GeV region. 
The simplest model, the quark-pair creation model (QPC)~\cite{martin_yaouanc}, allows for
quark pairs to be 
`popped'
out of the vacuum. This model by itself is insufficient
to describe the hadronic cross-section. Ono {\it et al.}~\cite{ono_qhm} 
are
able to reproduce the spectrum by introducing a $b\bar{b}g$ hybrid at 10683 MeV
(the QHM model).
A third model, the unitarized quark model (UQM)~\cite{tornqvist_byers} includes
coupled channel effects between $b\bar{b}$ states via real 
(two-meson) intermediate states. The QHM predicts
$\bbstar$ is dominant at 10865~MeV, with relative rates of 
$\bstbst$:$\bbstar$:$\bbbar$ $\sim$ 1:3:6. On the other hand, the 
UQM also reproduces the hadronic cross-section, but predicts that
$\bstbst$ is dominant, {\it e.g.,} $\bstbst$:$\bbstar$:$\bbbar$ $\sim$ 2:1:0.3.

A more thorough experimental study of the $\fives$ has been undertaken 
by CLEO, followed by Belle, to better understand the hadronic cross-section
in this region as well as assess the potential for a future high luminosity
$e^+e^-$ machine operating at the $\fives$. 

In 2003, CLEO collected 0.42~$\ifb$ of data at the $\fives$ (10871 MeV)
In 2005, Belle collected about 1.9~$\ifb$ on the $\fives$ resonance, followed
by a larger 
sample of 
22~$\ifb$ 
a year later. Here we report on results from 
CLEO and 
the
first Belle data set.

\section{HADRONIC RESONANCE CROSS SECTION}

The hadronic resonance cross section is measured by counting hadronic events 
recorded at the $\fives$ using various selection criteria employed to reduce the 
continuum background. The continuum background is estimated using data taken
below the $\fours$, and is scaled by: 
$({\cal{L}}_{\fives}/{\cal{L}}_{\fours})\times(s_{\fives})/s_{\fours})$, where
$ {\cal{L}}_i$ and $s_i$ are the 
integrated
luminosities and center-of-mass energies squared 
at the $\fours$ and $\fives$. The $\fives$ cross-sections are measured to be
$(0.301\pm0.002\pm0.039)$~nb~\cite{cleo5s_stone} and $(0.302\pm0.002\pm0.015)$~nb~\cite{drutskoy1}, 
by 
CLEO and Belle, respectively.

\section{$B$ AND $B^0_s$ OVERVIEW}

Many of the same techniques of exclusive reconstruction that are 
employed at the $\fours$ are used
to reconstruct $B$ and $B^0_s$ mesons at the $\fives$. 
The kinematic variables $\de=\ebeam-E_B$ and the beam-energy constrained
mass, $\mbc=\sqrt{\ebeam^2-p_B^2}$ are used.
Here, $\ebeam$ is the beam energy
and $E_B$ ($p_B$) is the energy (momentum) of the $B$ candidate. 
Because of the proximity of the $B^0_s$ mass to the beam energy,
$\bsbsbar$, $\bsbsstar$, and $\bsstbsst$ are kinematically well-separated. 
For $B$ mesons, $\bbbar$, $\bbstar$, $\bstbst$
and $\bbbarpipi$ are also kinematically separated. 
$\bbbarpi$, $\bbstarpi$ 
and
$\bstbstpi$ occupy a broad, but distinct region in
$\de-\mbc$ and overlap with one another, and partially with $\bbbarpipi$.
Because $B_{(s)}$ candidates are reconstructed
(not $B_{(s)}^*$), only
for $\bbbar$ and $\bsbsbar$ does $\de$ peak at zero and $\mbc$ peak at the
physical $B$ and $B^0_s$ masses. For the other processes, $\de$
and $\mbc$ are offset by a predictable amount. By measuring the yields in
each of these kinematically defined regions, one can determine the $B$ and
$B^0_s$ fractions at the $\fives$ as well as the 
contributing
subprocesses.

A second approach for obtaining the $B$ and $B^0_s$ yields at the $\fives$
is to use inclusive particle yields. This technique exploits
the large difference in inclusive rates for $B$ and $B^0_s$ to various
particles, such as $D_s$, $D^0$, and $\phi$.

\section{EXCLUSIVE $B$ MESON FINAL STATES AT THE $\fives$}

CLEO reconstructs $B$ mesons via their decays $B\to\jpsi (K^+, K^0_S, K^{*0})$
and $B\to D^{(*)}(\pi,\rho)$ (25 channels in total)~\cite{b5s_cleo}. First,
the inclusive $B$ yield is determined by computing the 
invariant 
mass of
all $B$ meson candidates with $-200< \de < 450$~MeV and $5272<\mbc<5448$~MeV.
Unlike $\mbc$, invariant mass does not depend on the particular $\fives$ decay
mode. A fit to the invariant mass spectrum yields a signal of $(53.2\pm9.0)$ events.
Using efficiencies ($\epsilon$) from Monte Carlo (MC) simulation and branching 
fractions (${\cal{B}}$) from the PDG~\cite{pdg}, 
it is found $\sum{\epsilon_i {\cal{B}}_i}=7.2\times10^{-4}$, 
where the sum is over the
25 decay modes. 
From this it is concluded that
$\sigma(e^+e^-\to\bbbar (X))=(0.177\pm0.030\pm0.016)$~nb. 
Using this cross-section and the measured total hadronic cross-section,
the $B^0_s$ fraction in $\fives$ decays, $f_s$, is computed to be
$f_s = 1-(\sigma_{\bbbar X}/\sigma_{\rm had})=(0.41\pm0.10\pm0.09)$.
To determine the contributing subprocesses, 
events are selected in the expected signal
band in the $\de-\mbc$ plane and projected onto the $\mbc$ axis. The
resulting distribution of events is shown in Fig.~\ref{fig:mbc_b}. A
dominant $\bstbst$ peak is observed, along with a smaller contribution
from $\bbstar$, and an insignificant yield from $\bbbar$. The distributions
are fit to the sum of 3 Gaussians, with resolutions determined from MC
and the mass difference between the $\bstbst$ and $\bbbar$ peaks 
constrained to the expected value of 47.5~MeV. The fitted yields are
$3.7^{+3.1}_{-2.4}$ $\bbbar$, $10.3\pm3.9$ $\bbstar$ and $31.4\pm6.1$ $\bstbst$
events. Insignificant yields in $B^{(*)}\bar{B}^{(*)}\pi$ and $\bbbarpipi$
lead to 90\% confidence level upper limits of 13.1 and 6.4 events, respectively.
Thus, it is found 
that $(74\pm15\pm8)\%$ of the $B$ production is through $\bstbst$
and $(24\pm9\pm3)$ is from $\bbstar$. For $\bbbar$, $B^{(*)}\bar{B}^{(*)}\pi$
and $\bbbarpipi$ 
90\% upper limits of 22\%, 32\% and 14\%, respectively, are obtained. 
These results strongly disfavor the QHM and are consistent
with the UQM. 

     The $\mbc$ peak position for $\bstbst$ and the corresponding
peak position from $\bsstbsst$~\cite{bs5s_ian} can be used to make a precise measurement of the
$B^*_s$ mass. Noting that $M(B_s^*)-M(B^*)=\mbc(\bsstbsst)-\mbc(\bstbst)+1.6$~MeV,
and using the well-measured $B^*$ mass~\cite{pdg}, 
gives 
$M(B_s^*)=(5411.7\pm1.6\pm0.6)$~MeV, which provides a significant improvement over
the previous measurement. The M1 mass splitting for $B^0_s$ is thus found to be
$(45.7\pm1.7\pm0.7)$~MeV, which is consistent with the value in the $B$
system of $(45.78\pm0.35)$~MeV~\cite{pdg}, 
as expected
from heavy quark symmetry.

\begin{figure}[bht]
  \includegraphics[height=.33\textheight]{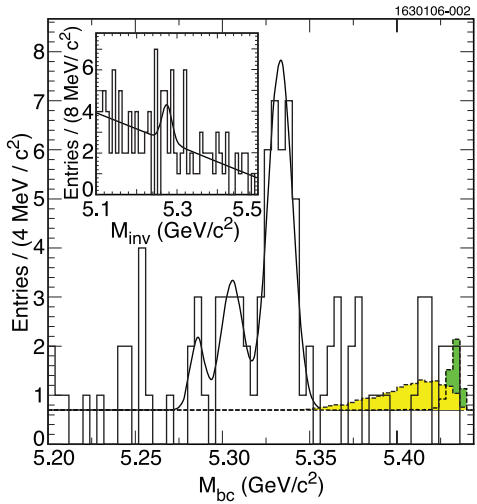}
  \vspace{-0.4in}
  \caption{$\mbc$ for $B$ signal candidates
in CLEO.
The histogram is the
data and the line is a fit as described in the text. The shaded regions
indicate the shapes expected from $B^{(*)}\bar{B}^{(*)}\pi$ and
$\bbbarpipi$. The inset shows the invariant mass for these particular
final states, as described in the text.
\label{fig:mbc_b}}
\end{figure}

\vspace{-0.25in}
\section{EXCLUSIVE $B^0_s$ MESON FINAL STATES AT THE $\fives$}

Both CLEO and Belle reconstruct $B^0_s\to\jpsi(\phi,\eta,\eta^{\prime})$~\cite{cleo_only}
and $B^0_s\to D_s^{(*)}(\pi,\rho)$ candidates. In Fig.~\ref{fig:bs-excl} we
show the $\de$ versus $\mbc$ and $\mbc$ distribution for 
$B^0_s\to D_s^{(*)}(\pi,\rho)$ candidates from the CLEO $\fives$ data
sample. The three boxes, from left to right, indicate signal regions
for $\bsbsbar$, $\bsbsstar$, and $\bsstbsst$. All 10 
signal candidates 
are $\bsstbsst$. An additional 
4 events are observed in the $\jpsi$ modes, all of which are
also $\bsstbsst$. Using efficiencies from simulation and branching
fractions from the PDG~\cite{pdg}, 
yields 
$f_s^*=\sigma(\fives\to\bsstbsst)/\sigma(\fives)=(37^{+14}_{-12}\pm9)\%$,
where $f_s^*$ neglects the evidently small $\bsbsbar$ and $\bsbsstar$ 
contributions. Belle observes a total of
23 events (21 $\bsstbsst$, 2 $\bsbsstar$, 0 $\bsbsbar$)
and determines $94^{+6}_{-9}\%$ of $B^0_s$ are from
$\bsstbsst$~\cite{drutskoy2}.

\begin{figure}[bht]
  \centering
  \includegraphics[height=.21\textheight]{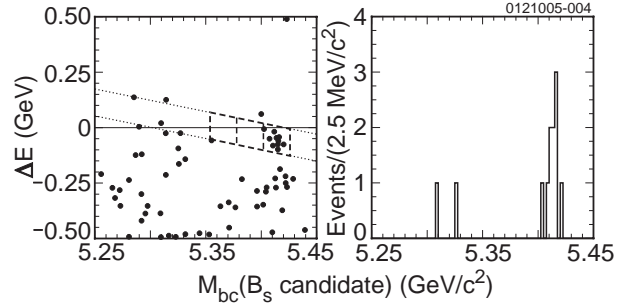}
  \vspace{-0.6in}
  \caption{$\de$ versus $\mbc$ (left) and $\mbc$ (right) for 
    $B^0_s\to D_s^{(*)}(\pi,\rho)$ candidates in the CLEO $\fives$ data 
    sample.
The superimposed boxes are discussed in the text.
\label{fig:bs-excl}}
\end{figure}

\section{INCLUSIVE ANALYSES}

Inclusive analyses are senstive to $f_s$, provided the {\it probe} particle
chosen has sufficiently different production rate 
for
$B$ and $B^0_s$
decays. CLEO has used both $D_s$ and $\phi$ mesons, both of which are
expected to have substantially higher rates from $B^0_s$ than $B$ 
mesons~\cite{cleo5s_stone}. Belle followed up with a similar pair
of analyses that use $D_s$ and $D^0$ mesons~\cite{drutskoy2}.
Measurement of the $B^0_s$ fraction, $f_s$, is obtained
using an expression of the form:

\begin{equation} 
\begin{split}
{\cal{B}}(\fives\to D_s X)=N_{\fives}(2f_s{\cal{B}}(B^0_s\to D_s X)+\\
(1-f_s){\cal{B}}(\fours\to D_s X)),
\end{split}
\label{eq:eq1}
\end{equation}

\noindent where $N_{\fives}$ is the number of $\fives$ decays, and
we have inserted $D_s$ as the probe particle as an example.
Here one measures ${\cal{B}}(\fives\to D_s X)$ and ${\cal{B}}(\fours\to D_s X)$,
and makes a model-dependent estimate of 
${\cal{B}}(B^0_s\to D_s X)=(92\pm11)\%$~\cite{cleo5s_stone}, leaving only $f_s$ unknown.
Using similar arguments, it is 
estimated that ${\cal{B}}(B^0_s\to D^0 X)=(8\pm7)\%$~\cite{drutskoy2}.
The first measurement of $f_s$ using this technique was reported by CLEO,
where they measure: $f_s = (16.8\pm2.6^{+6.7}_{-3.4})\%$~\cite{cleo5s_stone}
using $D_s\to\phi\pi^+$ mesons. 
The $D^0$ and $D_s$ scaled momentum spectra from Belle using 
$\fives$ data (points) and the below-$\fours$ continuum (histogram)
are shown in Fig.~\ref{fig:dsd0inc}.
The excess above continuum corresponds to the left-hand side of Eqn.~\ref{eq:eq1}.
Using ${\cal{B}}(B\to D_s X)=(8.7\pm1.2)\%$ and 
${\cal{B}}(B\to D^0 X)=(64.0\pm3.0)\%$~\cite{pdg}, Belle obtains:
$f_s = (17.9\pm1.4\pm4.1)\%$ from $D_s$ and $f_s = (18.1\pm3.6\pm7.5)\%$ from $D^0$,
which are consistent with each other and with the previous CLEO result.

\begin{figure}[bht]
  \centering
  \includegraphics[height=.20\textheight]{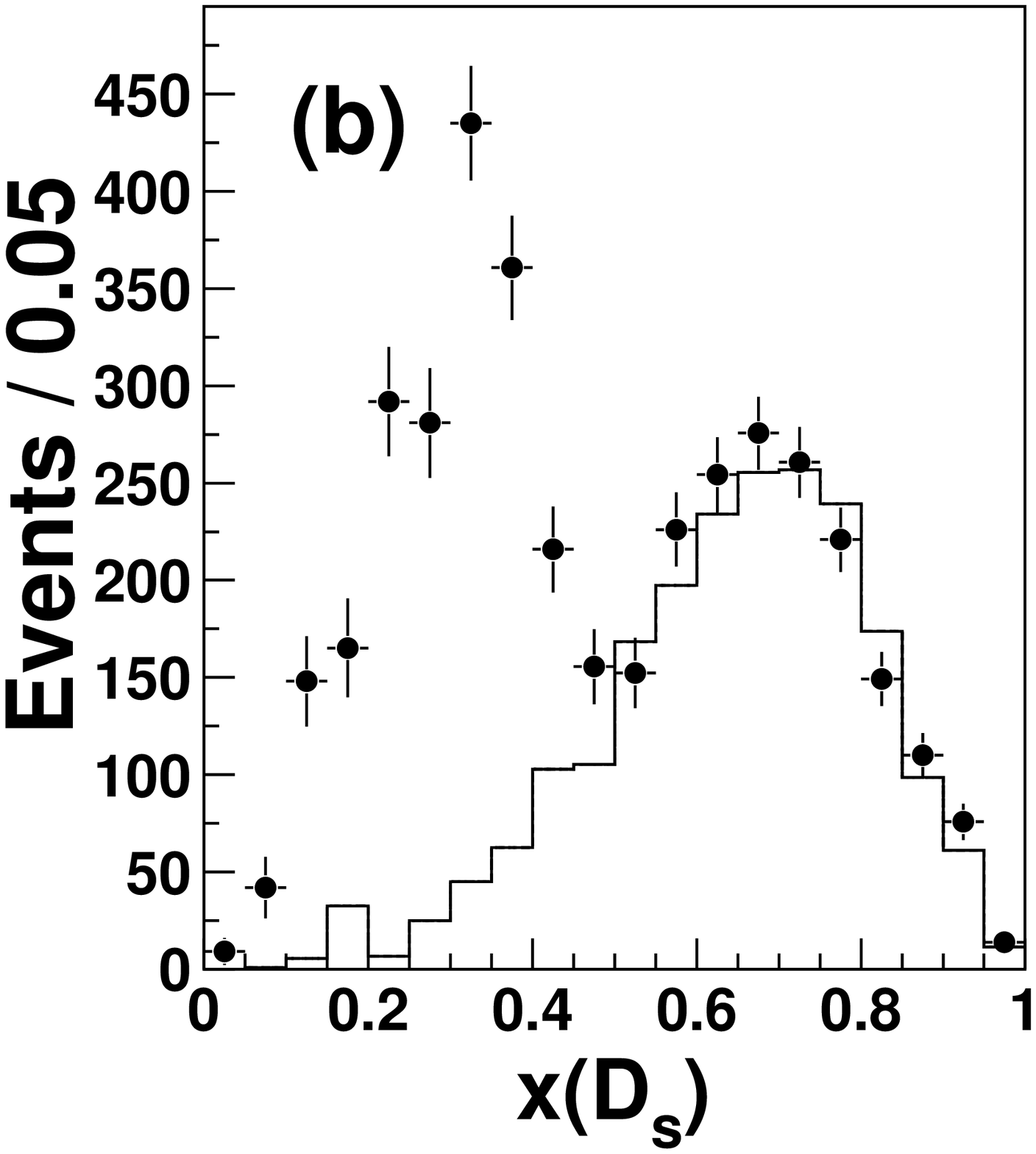}
  \includegraphics[height=.20\textheight]{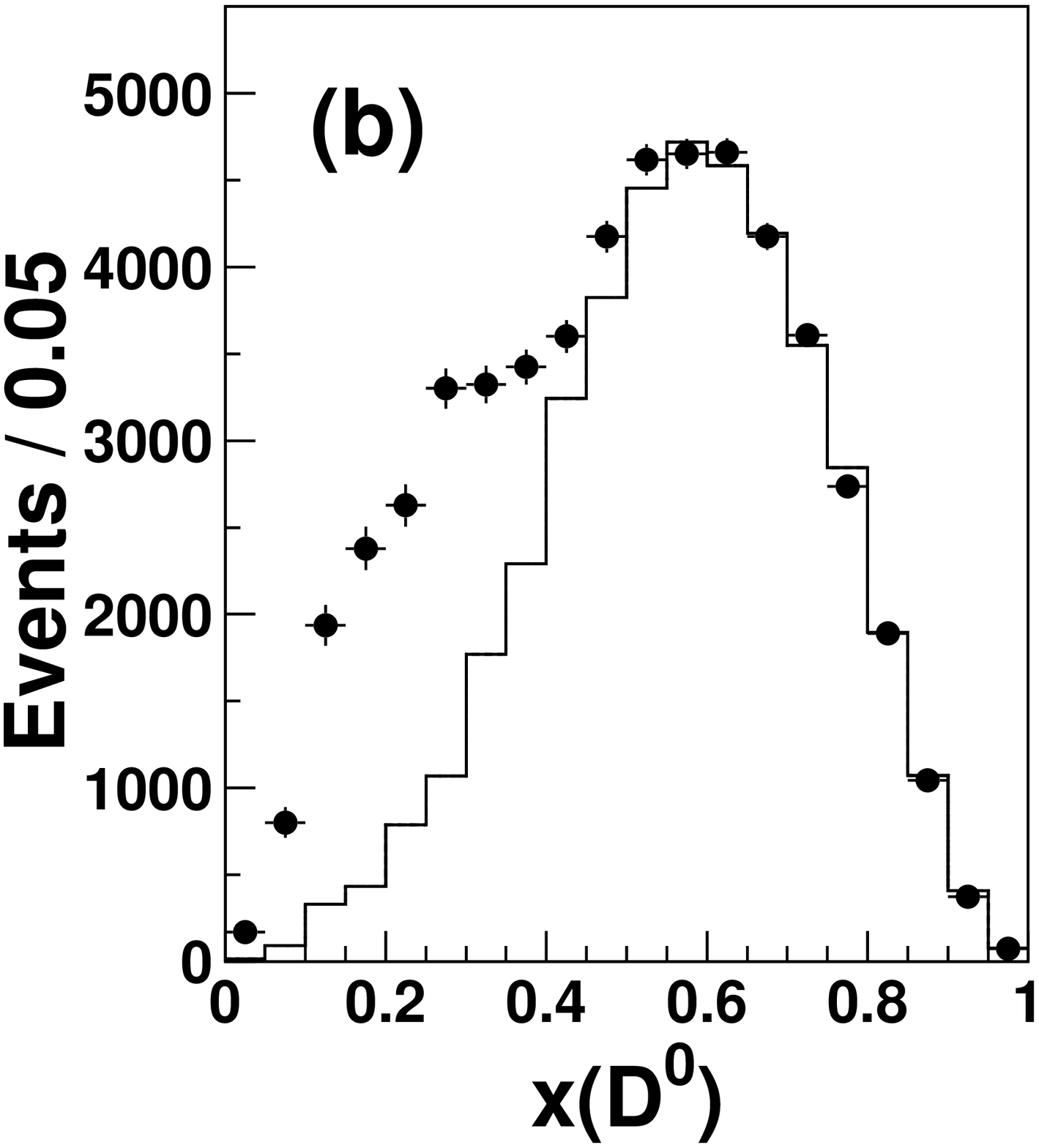}
  \vspace{-0.4in}
  \caption{Scaled momentum distributions, $x=p_D/\ebeam$, 
for (a)~$D_s^+\to\phi\pi^+$ and (b)~$D^0\to K^-\pi^+$ 
from Belle. The points are data taken at the $\fives$ resonance and the histogram
is the continuum obtained from $\fours$ data, normalized by the ratio of luminosities 
and $s=E_{\rm cm}^2$.     
\label{fig:dsd0inc}}
\end{figure}

CLEO has also performed an analysis using inclusive $\phi$ production~\cite{cleo5s_stone},
where $\phi\to K^+K^-$. The measured yields as a function of scaled momentum are
shown in Fig.~\ref{fig:phi-5s}. The first bin $0<x<0.05$ is obtained from MC simulation,
because in this bin the kaons generally have too low momentum to be reconstructed. The
uncertainty is taken to be 100\% of its value.
From the measured $D^0\to\phi X$, $D^+\to\phi X$ and $D_s\to\phi X$
branching fractions~\cite{cleo_incd}, 
it is deduced that
most $\phi$ mesons in $B_{(s)}$ decay come from the cascade $B_{(s)}\to D\to\phi$ or 
$B_{(s)}\to D_s\to\phi$, which allows one to make a model-dependent estimate 
${\cal{B}}(B^0_s\to\phi X)=(16.1\pm2.4)\%$. One then uses Eqn.~\ref{eq:eq1} to obtain
$f_s = (24.6\pm2.9^{+11.0}_{-5.0})\%$.

The combined CLEO average, using exclusive $B$ mesons, and inclusive $D_s$ and $\phi$
mesons, is $f_s=21^{+6}_{-3}\%$.

\begin{figure}
  \centering
  \includegraphics[height=.31\textheight]{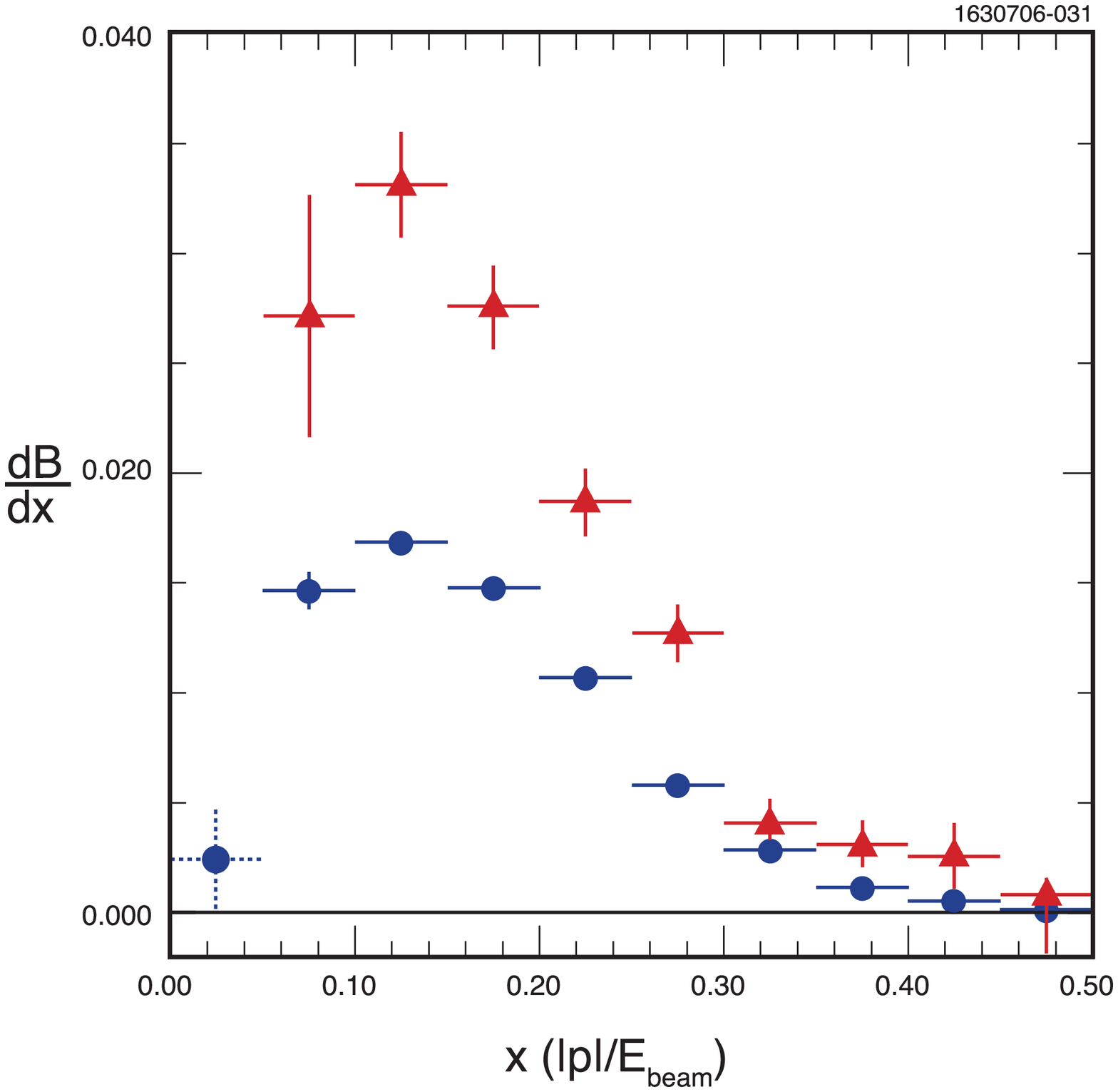}
  \vspace{-0.4in}
  \caption{
CLEO 
continuum-subtracted scaled momentum spectra for $\fives\to\phi+X$ 
(filled 
circles) and
$\fours\to\phi+X$ (triangles).
\label{fig:phi-5s}}
\end{figure}

These studies firmly establish the $B^0_s$ production rate to be about 25\%,
with roughly 90\% of it produced as $B^{0*}_s\bar{B}^{0*}_s$.
It is also found that 
$\bstbst$ is dominant, with $\bbstar$ about (1/3)$\bstbst$. Other processes
are less than about 25\% of $\bstbst$. These results are consistent with
the UQM, and inconsistent with the QHM. 

\section{$\Delta\Gamma_s$ AND RARE DECAYS}

With a large sample of data collected on the $\fives$, one can begin to search for
rare $B^0_s$ decays. While many of the decays can be searched for in $p\bar{p}$ (CDF, D0)
and $pp$ (LHCb at CERN, starting 2008), the $e^+e^-$ environment enjoys low
particle multiplicity and good signal-to-background for low energy photons.
Final states with soft photons are particularly difficult for
detectors at hadron machines, and so this is an area 
where 
machines operating at the
$\fives$ can make significant contributions. Taking 
$\sigma(e^+e^-\to B^0_s\bar{B}^0_s)=0.075$~nb (25\% of the resonance cross-section), a
50~$\ifb$ dataset, which could be collected by Belle in a couple of months, 
yields about 7.5 million $B^0_s$ decays. 

In the Standard Model, the width difference, $\Delta\Gamma^{CP}_s$ between the CP 
eigenstates, $B_s^{odd}$ and $B_s^{even}$ is given by  $\Delta\Gamma^{CP}_s=2|\Gamma_{12}|$,
where $\Gamma_{12}$ is the off-diagonal matrix element in the $B^0_s$ decay matrix,
which connects $B^0_s$ and $\bar{B}^0_s$ via on-shell intermediate states. These
matrix elements are saturated by tree-diagrams, such as $B^0_s\to D_s^{+(*)}D_s^{-(*)}$,
and therefore $\Gamma^{CP}_s$ is expected to be insensitive to new physics. Because
there are more CP 
even 
than CP odd final states in $B^0_s$ decays, one expects
$\Delta\Gamma^{CP}_s\simeq(0.12\pm0.06)$\cite{nierste}. In the presence of CP
violation, one finds a width difference between the mass eigenstates of 
$\Delta\Gamma_s=\Delta\Gamma^{CP}_s\cos\phi$, where $\phi=arg(-M_{12}/\Gamma_{12})$,
is the difference between the $B^0_s$ mixing phase and the phase of $-\Gamma_{12}$.
In the Standard Model, $\phi$ is expected to be small, of order 0.01. 
Therefore, one expects $\Delta\Gamma^{CP}_s\simeq\Delta\Gamma_s$. Since new
physics could affect $B^0_s$ mixing, {\it i.e.,} the phase of $M_{12}$, new physics
could drive $\Delta\Gamma_s$ away from $\Delta\Gamma^{CP}_s$. Precise measurements
of $\Delta\Gamma^{CP}_s$ can be obtained by measuring the branching fractions
and CP content of final states accessible to $B^0_s$ and $\bar{B}^0_s$.
Such final states are dominated by $D_s^{+(*)}D_s^{-(*)}$, but also
include $\jpsi\phi$, $\jpsi\eta$ $\jpsi\eta^{\prime}$, etc. Under certain
assumptions~\cite{nierste}, it can be shown that $B^0_s\to D_s^{+(*)}D_s^{-(*)}$ is pure
CP even, and neglecting the charmonium states,
one finds $\Delta\Gamma^{CP}_s/\Gamma_s\simeq 2{\cal{B}}(B^0_s\to D_s^{(*)+}D_s^{(*)-})$.
With a 50~$\ifb$ $\fives$ data sample, one would expect $\sim$10 events in each decay 
mode~\cite{drutskoy3}, yielding $\sim$25\% accuracy on ${\cal{B}}(B^0_s\to D_s^{+(*)}D_s^{-(*)})$.
If these assumptions are relaxed, one must measure the CP fraction
of the various vector-vector final states, which would require
substantially larger statistics.

Various rare loop-induced decay modes have been searched for in the 
Belle data sample, including $B^0_s\to\gamma\gamma$, $B^0_s\to\phi\gamma$, 
and $B^0_s\to K^+K^-$~\cite{drutskoy1}. The limit 
${\cal{B}}(B^0_s\to\gamma\gamma)<0.53\times10^{-4}$, is already more
stringent than the current world average.

\section{CONCLUSIONS}

In summary, we report on measurements of the $B^0_s$ fraction, $f_s$ at
CLEO and Belle, and find about 25\% of the $\fives$ decays produce $B^0_s$,
predominantly in the form of $\bsstbsst$. We also find a dominance of $\bstbst$
among ordinary $B$ decays. Both of these observations are consistent with the
UQM and inconsistent with the QHM. An $e^+e^-$ machine operating at the
$\fives$ produces about 15 million $B^0_s$ per 100~$\ifb$. Studies of decay
channels that produce soft photons could provide complementary information to
$B^0_s$ studies at hadron machines.

\end{document}